\documentclass[twocolumn,tightenlines,prd,floatfix,superscriptaddress]{revtex4}
\usepackage{epsfig} 
\usepackage{amssymb} 
\usepackage{amsmath} 
\usepackage{amsfonts}

\newcommand{\be}{\begin{equation}}
\newcommand{\ee}{\end{equation}}

\newcommand{\GeV}{{\rm GeV}}

\newcommand{\cm}{{\rm cm}}
\newcommand{\pb}{{\rm pb}}
\newcommand{\eff}{{\rm eff}}

\begin{document}

\title{ Muon Fluxes from Dark Matter Annihilation} 

\author{Arif Emre Erkoca} 
\affiliation{Department of Physics, University of Arizona, Tucson, AZ 85721}

\author{Mary Hall Reno}
\affiliation{Department of Physics and Astronomy, University of Iowa, Iowa City, IA}

\author{Ina Sarcevic}
\affiliation{Department of Physics, University of Arizona, Tucson, AZ 85721}
\affiliation{Department of Astronomy and Steward Observatory, 
University of Arizona, Tucson, AZ 85721}

\begin{abstract}
We calculate the muon flux from annihilation of the dark matter in the 
core of the Sun, in the core of the Earth and from cosmic diffuse neutrinos 
produced in dark matter annihilation in the halos.  
  We consider model-independent 
direct neutrino production and secondary neutrino production from the 
decay of taus produced in the annihilation of dark matter.  
We illustrate how muon energy distribution from dark matter annihilation 
has a very different shape than muon flux from atmospheric neutrinos.  
We consider both the upward muon flux, when muons are created in the 
rock below the detector, and the contained flux when muons are 
created in the (ice) detector.  
We contrast our results to the ones previously obtained in the literature, 
illustrating the importance of properly treating muon propagation and 
energy loss.  We comment on neutrino flavor dependence and their detection.

\end{abstract}

\maketitle

\section{Introduction}
The dark matter problem, where more matter is 
required to account for gravitational forces observed on astronomical objects 
than it is visible, has persisted for more than 
seven decades \cite{DM1}.    
Observations of galactic rotation 
curves \cite{rotation_curves}, 
orbital velocities of galaxies within clusters 
\cite{orbital_velocity}, 
anisotropies of the cosmic microwave 
background (CMB) \cite{CMB}, distance measurements from Type Ia 
supernovae 
(SN) and 
baryon acoustic oscillations (BAO) \cite{SN_BAO}, and large scale structure 
\cite{largescale} all
imply the existence of cold 
(non-relativistic) dark 
matter (CDM) with an abundance of $23\%$ of the total density of the Universe ($\Omega_{CDM}=0.233\pm0.013$). In addition, the combination of the CMB, SN
and BAO data predicts that only $4\%$ of the total density of the Universe can be attributed to the baryonic matter ($\Omega_{baryons}=0.0462\pm0.0015$). Thus, the 
particle content of the CDM { can not be explained} in the context of the standard model. 

In all extensions of the standard model, there are many candidates to account 
for CDM: weakly interacting massive particles (WIMPs), axions, superheavy dark matter
(WIMPZILLAs), and solitons (Q-balls) \cite{DM1,superheavy,cand1,cand2} to name a few. Among these 
possibilities, a WIMP of mass of order 100 GeV provides a natural explanation for the 
observed density of dark 
matter today \cite{barger,WIMP}. These WIMPs were
abundant and in thermal equilibrium in the early Universe and then eventually ``froze out'' 
due to the Hubble expansion.  

An interesting coincidence, independent of 
the dark 
matter issue, is that the 100 GeV scale is the characteristic scale of new physics 
beyond the standard model according to naturalness arguments 
\cite{naturalness}. 
Collider experiments such as the Large Hadron Collider (LHC) at CERN will explore this new scale physics 
in the 
near future \cite{LHC}. The 
detection and characterization
of dark matter particles is possible in these LHC searches.  
However, the LHC experiments will not be able to determine
detailed properties of these particles such as whether they are 
stable and what their 
couplings are to other particles. 

Apart from the colliders, there are two independent but complementary approaches
to search for dark matter: direct and indirect detection \cite{gelmini}, including
dark matter accumulation in the Earth and Sun \cite{DMcapture,roulet,gould} and in the galactic center \cite{gcann},
and subsequent annihilation to 
neutrinos \cite{earlydmnu,dmnu,perez,edsjo}. There are a number of direct
detection experiments \cite{IGEXetc,XENON,CDMS,DAMA}. 
Direct searches can provide valuable data on the dark matter's 
couplings to 
the standard model. They all look for energy deposition via nuclear recoils from 
WIMP scattering by using different target nuclei and detection strategies, and expect to 
observe the same WIMP mass and cross sections. Currently, the strongest 
upper bounds ($\sim 10^{-7}$pb) on the spin independent WIMP-nucleon cross 
section of a 
WIMP with mass $\sim100$ GeV 
come from the {\bf XENON} Dark Matter Search (XENON) \cite{XENON} and 
the {\bf C}ryogenic {\bf D}ark {\bf M}atter {\bf S}earch (CDMS) 
\cite{CDMS} experiments.  
So far, contrary to the null results of all other direct searches, the 
{\bf DA}rk {\bf MA}tter (DAMA) collaboration has 
observed an annual 
modulation in 
their data \cite{DAMA} which is claimed to be due to dark matter particles in the galactic halo.  
Lately, some models have also been proposed to account for that modulation 
signal \cite{arkani,DMmodels}.

On the other hand, indirect searches for WIMPs through their annihilation 
(or sometimes decay) into standard model degrees of freedom such as positrons, antiprotons or $\gamma$ rays, 
 which has been explored in several experiments 
\cite{FGSTetc,HEAT,PAMELA,ATIC,PPB-BETS} 
 and 
neutrinos with experiments such as 
{\bf A}ntarctic {\bf M}uon {\bf A}nd {\bf N}eutrino {\bf D}etector 
{\bf A}rray (AMANDA) 
\cite{AMANDA}, 
IceCUBE \cite{IceCube}, Cubic Kilometer Size ({\bf KM3}) {\bf Ne}utrino 
{\bf T}elescope (KM3NeT) 
\cite{KM3NeT}. Observations in the recent years such as the 
excess in the positron 
fraction reported by {\bf H}igh {\bf E}nergy {\bf A}ntimatter {\bf T}elescope (HEAT) \cite{HEAT}, 
the {\bf P}ayload of {\bf A}ntimatter {\bf M}atter {\bf E}xploration and {\bf L}ight-nuclei {\bf A}strophysics 
(PAMELA) \cite{PAMELA}, {\bf A}dvanced {\bf T}hin {\bf I}onization {\bf C}alorimeter (ATIC) 
\cite{ATIC} and {\bf P}olar {\bf P}atrol {\bf B}alloon and {\bf B}alloon borne {\bf E}lectron {\bf T}elescope 
with {\bf S}cintillating fibers (PPB-BETS) \cite{PPB-BETS}, 
an excess in microwave emission around the 
galactic center \cite{wmaphaze} (also called the ``WMAP Haze"), 
a bright $511$ keV gamma-ray line from the Galactic Center region from 
{\bf INTE}rnational {\bf G}amma {\bf R}ay {\bf A}strophysics {\bf L}aboratory 
(INTEGRAL) 
\cite{INTEGRAL}, 
 and 
an excess in the flux of 1-10 GeV diffuse galactic $\gamma$ rays from {\bf E}nergetic {\bf G}amma {\bf R}ay {\bf E}xperiment {\bf T}elescope (EGRET) 
data \cite{EGRET}
have made researchers more 
excited in their quest for dark matter.     
Theoretical studies of the 
indirect dark matter detection via neutrino signals has recently received a lot of attention  
\cite{theory,menon}.  
  
In this paper, our focus is on 
the muon energy distributions from $\nu_\mu+\bar{\nu}_\mu$ 
in neutrino telescopes due to annihilation of WIMPs 
which are captured in the core of the Earth (or the Sun) via 
gravitational interaction, or from annihilation of relic neutrinos \cite{beacom}. 
As a result of these annihilations, neutrinos are produced at energies of the order of the mass of the WIMP and they interact on their way to the 
detector producing an observable muon flux. There is an extensive literature on WIMP annihilation 
in the Earth's or Sun's core \cite{cand1,DMcapture,perez,edsjo}.
Here, we present a systematic way of calculating this muon flux for a few choices of annihilation channel. Our results can be used to determine the muon flux as a function of energy 
for a specific dark matter model by summing all the contributions from each annihilation 
mode weighted with corresponding branching fractions.          
We also compare our muon energy distributions
with those obtained using other theoretical frameworks widely used in the literature 
\cite{barger,WIMP,perez,menon,edsjo}.  

The muon neutrino flux from weakly interacting dark matter annihilation ($\chi\chi$ annihilation)
depends on the dark matter capture rate and the dark matter annihilation rate. In the next section, we review the standard evaluation of the (muon)
neutrino flux. This is followed by a discussion of the theoretical framework of muon survival probabilities and the resulting muon flux. Results are shown in Section IV, followed by our conclusions in Section V. The Appendix includes details of the muon neutrino energy distribution from various decay modes of fermions $F$ in $\chi\chi\rightarrow F\bar{F}$ where $F=\nu,\ \tau,\ c$ and $b$.

\section{Neutrino flux from dark matter annihilation}

The dark matter particles can be captured in the core of the Sun or the Earth by interacting with the nuclei in the medium. 
This results in a WIMP density in the core that is considerably higher than in the galactic halo. 
The capture rate ($C$) depends on the composition of the medium, the 
WIMP-nucleus 
interaction cross sections ($\sigma^i_0$), the WIMP mass ($m_\chi$), the local dark matter density 
($\rho^\chi$) and 
velocity 
($\overline{v}$)
distribution 
of the WIMPs in the halo. After being accumulated in the core of these dense objects, 
the WIMPs annihilate with rate $\Gamma_A$ into standard model 
particles which may further 
decay into neutrinos. These neutrinos can reach Earth-based detectors and create fluxes of charged leptons as a consequence of neutrino
charged-current (CC) interactions. 

The resulting fluxes depend on how the capture and annihilation processes have occurred initially, 
however, in 
equilibrium these two processes are related: for every two WIMPs captured, 
one annihilation takes place so $\Gamma_A=C/2$. This 
equilibrium condition leads to a maximal flux which 
depends on the capture rate given by \cite{gould,cand1,roulet},    
\begin{eqnarray}\label{capture}
C &=& c\frac{\rho^\chi_{0.3}}{(m_\chi/\GeV)\overline{v}_{270}}\sum_i F_i(m_\chi)\,f_i\,\phi_i\, S(m_\chi/m_{N_i})\nonumber\\
 & &\times\frac{\sigma^i_0}{10^{-8}\ \pb}\frac{1\  
\GeV}{m_{N_i}},
\end{eqnarray}
where
\begin{equation}
\rho^\chi_{0.3}=\frac{\rho^\chi}{0.3\ \GeV/\cm^3}\,,\;\;\;  \;\;\; \overline{v}_{270}=\frac{\overline{v}}{270\ {\rm km/s}}
\end{equation}
and
\begin{equation}
c=\left\{
   \begin{array}{lr}
   4.8\times10^{11}{\rm s}^{-1}&  \mbox{Earth}, \\
   4.8\times10^{20}{\rm s}^{-1}&  \mbox{Sun}.
   \end{array}
   \right.
\end{equation}

The summation in Eq.\ (\ref{capture}) is over all species of nuclei 
in the astrophysical object, $m_{N_i}$ is the mass of the $i$th nuclear species with 
mass fraction $f_i$ relative to the Sun (or the Earth). The kinematic 
suppression factor, 
denoted by $S(m_\chi/m_{N_i})$, for a capture of WIMP of mass 
$m_\chi$ from a nucleus of mass $m_{N_i}$ is given by \cite{gould,cand1,roulet}
\begin{equation}
S(x)=\left(\frac{A^{1.5}}{1+A^{1.5}}\right)^\frac{2}{3}
\end{equation}
where
\begin{equation}
A(x)=\frac{3}{2}\frac{x}{(x-1)^2}\left(\frac{<v_{esc}>}{\overline{v}}\right)^2. 
\end{equation}
For the Sun, $<v_{esc}>=1156\ {\rm km/ s}$ and for the Earth, $<v_{esc}>=13.2\ {\rm  km/ s}$.  We also note that $S(x)\rightarrow1$ for $x\rightarrow1$, which means 
that the capture process is kinematically suppressed if $m_\chi$ differs from $m_{N_i}$, and there is no suppression if these masses are the same. 

The other quantities 
in the capture rate expression are the form factor suppression $F_i(m_\chi)$ and the velocity distribution function $\phi_i$ of the $i$th element. The former one is due to the 
finite 
size 
of the nucleus which disrupts the coherence in the scattering process, thus, the form factor suppression is a negligible effect for capture from scattering with 
hydrogen and helium whereas it becomes important for larger nuclei. The velocity distribution function $\phi_i$ depends on the velocity distribution squared 
of the element averaged over the volume of the astrophysical object 
($<v^2_i>$) and is given as \cite{gould,cand1,roulet},
\begin{equation}        
\phi_i=\frac{<v^2_i>}{<v^2_{esc}>} .
\end{equation}

If the massive astrophysical object is far from equilibrium, 
 which is most likely the case for the Earth, the annihilation rate is not only 
dependent on the capture 
rate but also on the annihilation cross section ($\sigma$) via,
\begin{eqnarray} 
\Gamma_A &=& \frac{C}{2}\tanh^2(t_0\sqrt{CC_A})\nonumber\\
\end{eqnarray}
where 
\begin{eqnarray}
C_A &=& \frac{<\sigma {v}>}{V_{\eff}}
\end{eqnarray} 
and $V_{\eff}$ is the effective volume of the core of the Earth or Sun, while $t_0$ is the age of the 
solar system. It is obvious from this relation that 
the equilibrium condition holds only when $t_0\sqrt{CC_A}\gg1$.

The flux of neutrinos of flavor $i$ from dark matter annihilation into standard 
model particles can be written as
\begin{equation}\label{neutrino}
\left(\frac{d\phi_\nu}{dE_\nu}\right)_i=\frac{\Gamma_A}{4\pi{R^2}}\sum_F
B_F\left(\frac{dN_\nu}{dE_\nu}\right)_{F,i} ,
\end{equation}
where $(dN_\nu/dE_\nu)_{F,i}$ is the differential energy spectrum of
neutrino flavor
$i$ from { production} of particles in channel $F$. In general, this energy 
spectrum is a function of the neutrino energy $E_\nu$ and 
the energy
of the produced particle, $E_{in}$.  
The differential neutrino energy spectra from a few dark 
matter annihilation channels are given in the
Appendix.  The quantity
$R$ is the
Sun-Earth distance for neutrinos produced in the core of the Sun,
or the radius of the Earth for the neutrinos created in the core of the Earth.  
The sum in Eq. (\ref{neutrino}) 
is over all annihilation channels $F$ weighted with corresponding branching fractions 
$B_F$. 

Neutrinos can be detected via their charged current interactions near or in the detector. To avoid the downward muon background, upward events where neutrinos interact with the 
nucleons in the rock below the detector producing muons  which then travel up through the detector
are one category of events. The other are contained events, in which the muon neutrinos produce muons  
in the detector ice.   
In the following 
sections, we focus on evaluating muon energy distribution from 
interactions of the neutrinos produced in the annihilation of the 
dark matter in the core of the Earth, in the core of the Sun and 
cosmic diffuse neutrinos from dark matter annihilation in the halos.  

\section{Muon Flux}

The muon flux from muon neutrinos from DM annihilation depends on the flux of muon neutrinos
as calculated with Eq. (\ref{neutrino}) and attenuation, tau neutrino regeneration and 
neutrino mixing in transit to the detector. 
For the energies of interest, neutrino mixing for DM annihilation in the Earth's core 
is not important.  For annihilations in the Sun, neutrino mixing and tau neutrino 
regeneration may affect the flux of muon neutrinos when $m_\chi$ is large.  
  We neglect these effects because we consider the value of $m_\chi$ for which 
  only moderate modification of the muon neutrino flux is expected \cite{barger}.  

For upward events where the muon is produced outside the detector, muon energy 
loss is important. The most straightforward evaluation is for contained events, 
so we start with this case.

In the sections below, we focus on the neutrino induced flux of muons, produced either in the detector or near the detector with muon energy loss included.
The detector may be modeled by an effective area. For IceCube, an effective area for muons, $A_{eff}(E_\mu,\theta)$, can be simply parameterized as a function of the  muon energy at the detector \cite{gghm}. 

\subsection{Contained Events}

Contained events involve neutrino conversions within the detector volume. Denote the muon neutrino flux from a source of DM annihilations in the Earth's core or the Sun's core
at location $R$ from what is effectively a point source by
$$ \frac{d\phi_\nu}{dE_\nu}(E_\nu,R)\ .$$
The probability of the conversion of a neutrino with energy $E_\nu$ into a 
muon with energy $E_\mu$ over
a distance $dr$ through CC interactions is given by
\begin{equation}
\label{conversion}
dP_{CC} = dr\, dE_\mu \left(\rho_p\frac{d\sigma^p_\nu(E_\nu,E_\mu)}{dE_\mu}+(p\rightarrow{n})\right)\ .
\end{equation}
Here, $\rho_p$ and $\rho_n$ are the number densities of protons and neutrons in the medium, respectively. We assume that $\rho_p = \rho_n = \frac{1}{2}N_A\rho$ where $N_A\simeq6\times10^{23}$ is Avogadro's number.  The differential cross sections
$d\sigma_{\nu}^{p,n}/dE_\mu$ are the weak scattering cross sections of (anti-)neutrinos on nucleus,
which can be approximated by \cite{barger}
\begin{equation}
\frac{d\sigma^{p,n}_\nu}{dE_\mu}=\frac{2m_pG^2_F}{\pi}\left(a^{p,n}_\nu+b^{p,n}_\nu\frac{E^2_\mu}{E^2_\nu}\right)
\end{equation}
with $a^{n,p}_\nu=0.25,0.15$, $b^{n,p}_\nu=0.06,0.04$ and $a^{n,p}_{\overline{\nu}}=b^{p,n}_\nu$, $b^{n,p}_{\overline{\nu}}=a^{p,n}_\nu$.
The contained event rate, for a detector with { size} $\ell$, is
\begin{eqnarray}
\label{eq:contained}
\nonumber
\frac{d\phi_\mu}{dE_\mu}&=& \int^{R+\ell}_R dr \int^{m_\chi}_{E_\mu}dE_\nu 
\frac{dP_{CC}}{dr\,dE_\mu}
\frac{d\phi_\nu}{dE_\nu}(E_\nu,R)\\
&+& (\nu\rightarrow \bar{\nu})
\end{eqnarray}
where the neutrino flux is essentially independent of position in the detector given the scale of the Earth.

The neutrino flux from DM annihilation  in the Earth's core is not attenuated, 
to a good approximation, until the neutrino interaction length approaches the radius of the Earth.  
This occurs at a neutrino energy of approximately 100 TeV. 
The neutrino flux 
in Eq. (\ref{eq:contained}) is given by Eq. (\ref{neutrino}), with $R\equiv R_E \simeq 6400$ km, 
the 
radius of the Earth, i.e. 
\begin{equation}\label{neutrinomu}
\frac{d\phi_\nu}{dE_\nu}(E_\nu, R_E)=\frac{\Gamma_A}{4\pi{R_E^2}}\sum_F
B_F\left(\frac{dN_\nu}{dE_\nu}\right)_{F,\mu}.  
\end{equation}

The density of the Sun is such that one needs to take into 
account neutrino attenuation due to its charged-current interactions as they 
pass through the 
Sun.  In our calculations, we approximate attenuation with the 
exponential decrease in the flux over a distance 
$\delta{r'}$ in the Sun, assuming that 
the composition of the Sun is mostly elemental hydrogen which has mass $m_H=0.931$ GeV.      
\begin{equation}
\frac{d\phi_\nu}{dE_\nu}(r'+\delta{r'})= 
exp(-\rho(r')\sigma_{CC}\delta{r'}/m_H)\frac{d\phi_\nu}{dE_\nu}(r')
\end{equation}
where 
the neutrino flux ${d\phi_\nu(r')}/{dE_\nu}$ is given by Eq. (12) with 
$R_E$ being replaced by the 
distance from the Sun to the Earth, $R_{SE}$.  
We use the Sun density profile given by \cite{density}
\begin{equation}
\rho(r')=236.93\, {\rm g/cm}^3\times\exp(-10.098\frac{r'}{R_{S}}), 
\end{equation}
where $R_S$ is the radius of the Sun and we sum up all $\delta{r'}$ contributions 
until neutrinos reach the surface of the Sun.  

\subsection{Upward Events and Muon Energy Loss}

High energy muons produced in neutrino charged-current interactions 
lose energy before they reach the detector as they travel through the rock or ice. 
The average energy loss of the muons with 
energy $E$ over a distance $dz$ during their passage 
through a medium with density $\rho$ is given by,  
\begin{equation}
\label{eq:dedx}
\langle \frac{dE}{dz}\rangle = -(\alpha +\beta E)\rho ,
\end{equation}
\noindent
where,  $\alpha \simeq 2\times 10^{-3}$ GeV\,cm$^2$/g accounts for the ionization energy 
loss and $\beta \simeq 3.0\times 10^{-6}$ cm$^2$/g accounts for the
bremsstrahlung, pair production and photonuclear interactions. The parameter $\alpha$ is relatively
insensitive to the composition of the material. The quantity $\beta$ depends on 
composition of the medium and varies slowly 
with energy \cite{groom,ls,range}, and the average energy loss 
formula is not strictly applicable because of stochastic 
energy losses \cite{ls,bb}. For our purposes here, given the other uncertainties, 
using a constant $\beta$ and approximating

\begin{equation}
\frac{dE}{dz}\simeq \langle \frac{dE}{dz}\rangle
\end{equation}
is sufficient.

With this assumption 
the initial energy at $z=0$, $E^i_\mu$, 
is related to the final energy $E^f_\mu$ after traveling
a distance $z$ by
\begin{equation}
\label{eq:emui}
E^i_\mu(z) = 
e^{\beta \rho z}E_\mu^f +(e^{\beta\rho z}-1)\frac{\alpha}{\beta}.  
\end{equation}
{ At low energies, for $E_\mu \leq 200$ GeV, the contribution from $\beta$ term is 
small (about $10-20\%$) and in this energy range,} 
\begin{equation}
E^i_\mu(z) 
\simeq E^f_\mu+\alpha\rho z\ .  
\end{equation}
Muons with energies of a few 100 GeV are stopped in the rock 
($\rho\simeq 2.6\ {\rm g/cm}^3$) before they decay.
As an example, the stopping distance for 500 GeV muons is roughly 1 km 
whereas the decay length, $\gamma c\tau$, for these muons
turns out to be about 3000 km. For 50 GeV muons, the decay length is about 
300 km, compared to a stopping distance of 100 m. 

The decay length information can still be 
included in the calculation by introducing the survival probability as the solution to the equation,    
\begin{equation}
\frac{dP_{\rm surv}}{dE_\mu} = \frac{P_{\rm surv}}{E_\mu c\tau\rho(\alpha+\beta E_\mu)/m_\mu}
\ .
\label{eq:surv}
\end{equation}
This leads us to the survival probability for a muon with initial energy $E^i_\mu$ and 
final energy $E^f_\mu$,  
\begin{equation}\label{survival}
P_{\rm surv}(E^i_\mu,E^f_\mu) = \Biggl( \frac{E^f_\mu}{E^i_\mu}\Biggr)^\Gamma
\Biggl(\frac{\alpha+\beta E_\mu^i}{\alpha+\beta E_\mu^f}\Biggr)^\Gamma
\end{equation}
where $\Gamma \equiv m_\mu/(c\tau \alpha\rho)$. 

With a distinction made between the energy of the muon when it is produced and the energy of the muon when it arrives at the detector for upward events, the formula for the upward muon flux is more complicated than Eq. (\ref{eq:contained}). Instead, we have
\begin{eqnarray}
\label{eq:upward}
\frac{d\phi_\mu}{dE_\mu}&=& \int^{R}_{R_{min}} dr \int^{m_\chi}_{E_\nu^{min}}dE_\nu 
\frac{dP_{CC}}{dr\,dE_\mu^i}\nonumber\\
&\times &
\frac{d\phi_\nu}{dE_\nu} P_{\rm surv}(E_\mu^i,E_\mu)\frac{dE_\mu^i}{dE_\mu}\nonumber\\
&+&(\nu\rightarrow \bar{\nu})\ .
\end{eqnarray}
where $E_\mu \equiv{E^f_\mu}$.  
The minimum neutrino energy in the integral is $E_\nu^{min} = E_\mu^i(z)$, 
where $z= R - r$.  The 
 maximum distance that muon travels to the detector and ends up with the 
 final 
energy 
 is  
 $R - R_{min}$ and the relationship between $E_\mu$ and $E^i_\mu$ given by 
Eq. (16).  

Consider as a specific example annihilation at the core of the Earth. The detector is at a distance 
$R=R_E\simeq 6400$ km from the core. In principle, $R_{min}=0$, however, only muons produced near the detector will have sufficient energies to make it to the detector with an energy above the detector threshold energy $E_{th}$. The muon average range is 
\begin{equation}\label{range}
R_\mu(E_\mu^i,E_{th})=\frac{1}{\beta\rho}\ln\left(\frac{\alpha+\beta{E_\mu^i}}{\alpha+\beta{E_{th}}}\right)
\end{equation} 
following from Eq. (\ref{eq:emui}). For an initial muon energy of 1 TeV, the 
muon average range is  
{ $1$ km} for a muon threshold energy of $50$ GeV.
In practice, then, $R_{min}=R_E-\Delta$, { where $\Delta$ is, in 
general, less than 1
kilometer for the energies considered here}.  
The upper limit for muon range is obtained by setting 
 $E_\mu^i = m_\chi$, in which case 
$\Delta = R_\mu(m_\chi,E_\mu)$.

A change of variable from $r$ to $z=R_E-r$ yields a more familiar form of the integral for the muon flux from DM annihilations in the Earth's core, 
\begin{eqnarray}
\label{eq:earthflux}
\nonumber
\frac{d\phi_\mu}{dE_\mu }&=&\frac{\Gamma_A}{4\pi{R_E^2}}\int^{R_\mu(m_\chi,E_\mu)}_0 dz e^{\beta
\rho z} \int^{m_\chi}_{E^i_\mu}dE_\nu 
\left(\frac{dN_\nu}{dE_\nu}\right)_{F,\mu}\\ \nonumber
&\times &
\left(\frac{E_\mu}{E^i_\mu}\frac{\alpha +\beta E_\mu^i}
{\alpha + \beta E_\mu}\right)^\Gamma
\times  \left\{\frac{d\sigma^p_\nu}{dE^i_\mu}\rho_p+(p\rightarrow{n})\right\}\\
&+& (\nu\rightarrow \bar{\nu}). 
\end{eqnarray}
Throughout Eq. (\ref{eq:earthflux}), the initial muon energy is implicitly a function of the final muon energy and the distance traveled, $E_\mu^i=E_\mu^i(E_\mu,z)$ from Eq. (\ref{eq:emui}).

Muon flux from DM annihilation in the Sun is given by 

\begin{eqnarray}
\label{eq:sunflux}
\nonumber
\frac{d\phi_\mu}{dE_\mu }&=&\frac{\Gamma_A}{4\pi{R_{SE}^2}}\int^{R_\mu(m_\chi,E_\mu)}_0 dz e^{\beta
\rho z} \int^{m_\chi}_{E^i_\mu}dE_\nu 
\left(\frac{dN_\nu}{dE_\nu}\right)_{F\mu}\\ \nonumber
&\times &
\left(\frac{E_\mu}{E^i_\mu}\frac{\alpha +\beta E_\mu^i}
{\alpha + \beta E_\mu}\right)^\Gamma
\times  \left\{\frac{d\sigma^p_\nu}{dE^i_\mu}\rho_p+(p\rightarrow{n})\right\}\\
&\times & 
\prod_{\delta{r'}} exp(-\rho(r')\sigma_{CC}\delta{r'}/m_H)\frac{d\phi_\nu}{dE_\nu}(r')\nonumber\\ 
&+& (\nu\rightarrow \bar{\nu}). 
\end{eqnarray}

\section{Results}

\subsection{DM annihilation in the Earth's core}

To illustrate the muon flux's dependence on muon energy, we begin with DM annihilation in the 
Earth's core. In addition to making a choice for $m_\chi$, 
one must also make some assumptions about the cross section and main channel to produce neutrinos.
For all of the figures for DM annihilation, we use $\sigma_0^i\simeq10^{-8}N^4_i$ pb \cite{perez} and the standard 
composition of the Earth as reviewed in Ref. \cite{cand1}. 

The upper curves in Fig. \ref{fig:earth} show our results for $\chi \chi\rightarrow
\nu_\mu\bar{\nu}_\mu$ with $B_{\nu_\mu}=1$, for upward events (dot-dashed curve) and contained events (dashed
curve) for $m_\chi=500$ GeV.  The lower dot-dot-dashed and dot-dash-dashed curves in  Fig.
\ref{fig:earth} come from $\chi\chi\rightarrow \tau^+\tau^-$ with $B_\tau=1$, followed by $\tau\rightarrow
\nu_\tau \mu\bar{\nu}_\mu$ according to the energy distribution in the Appendix. We choose 
the tau channel as representative of the three body decays that 
also occurs in heavy flavor semileptonic decays.

For direct production of neutrinos, $d\phi_\nu/dE_\nu\propto \delta(m_\chi - E_\nu)$. The cross section for neutrino production of muons smears the distribution. For 
contained 
events, one sees the smeared distribution directly in Fig. \ref{fig:earth}. Upward events have the additional energy redistribution from muon energy loss in transit
that shifts the muon energy distribution to lower energies, enhancing the lower energy flux relative to the contained flux, despite the fact that the range is shorter than $\ell = 1$ km. In the cascade of 
$\tau \rightarrow \nu_\mu\rightarrow \mu$, shown with the lower curves, there is never a high energy peak and the upward events are always below the contained events for this value of $m_\chi$. Only for 
$m_\chi$ sufficiently higher than 1 TeV could the upward events be enhanced relative to the contained events.

As an indication of the atmospheric neutrino background, we also show upward (solid curve)
and contained events (dotted curve)
from a solid angle defined by a cone of half-angle $1^\circ$ around the upward vertical direction.
We use a simple parametrization for the flux of atmospheric
$\nu_{\mu}+\bar{\nu}_\mu$ (in units of $\GeV^{-1}{\rm km}^{-2}{\rm yr}^{-1}{\rm sr}^{-1}$) 
from Ref. \cite{honda}, 
\begin{eqnarray} 
\frac{d\phi_\nu}{dE_\nu d\Omega} &=& 
N_0{E_\nu}^{-\gamma-1}\nonumber\\
&\times &\left(\frac{a}{1+bE_\nu{cos\theta}}+\frac{c}{1+eE_\nu{cos\theta}}\right)\ .
\end{eqnarray}
where the values of the parameters $N_0$, $\gamma$ $a$, $b$, $c$ and $e$ are 
given in Table 1.  
{The approximate angular resolution 
of the IceCube detector is $\theta=1^\circ$, however, DM annihilation can occur at angles larger than $1^\circ$. In particular, for a 500 GeV neutralino, it has been shown \cite{edsjo}
that most of the annihilation occurs within an angle of $\theta \sim 2.7^\circ$. With 
this larger nadir angle, the solid angle for the atmospheric background is increased by a factor of
$\sim 8$.

The shape of the background atmospheric flux is very different from the 
signal of contained events for direct annihilation of DM into neutrinos.  
With the atmospheric contained events in the figure multiplied by a factor of 8,
the contained event rate would dominate the background only for the high energy peak.
Our sample calculation is for $B_\nu \sigma_0^i =10^{-8} N_i^4$ pb$^{-1}$, in which the 
capture and annihilation rates are in equilibrium.  
For the secondary neutrino production, from $\tau$ decay, one needs $B_\tau \sigma_i^0 
\sim 10^{-7} N_i^4$ pb$^{-1}$ for 
this channel to be comparable to the background.  Clearly measurements of the shape of the 
muon flux, both contained and upward, would be useful in searching for the dark matter 
signal.  

These values of the cross section $\sigma_0^i$ required for signals on the order of the atmospheric background are sufficient for the 
condition for the equilibrium between capture and annihilation in the Earth's core to be satisfied.  Even though only with 
significant enhancements of the 
capture rate (i.e. WIMP-nucleon cross section) or DM annihilation rate, the Earth might 
be a source of measurable rates for DM annihilation to neutrinos, it is a useful
demonstration of the energy dependence of the muon flux.  
 
\begin{table}[t]
\begin{tabular}{|c|c|}
\hline
\hline
$\gamma $& 1.74\\
$a$ & 0.018\\
$b$ & 0.024\ GeV$^{-1}$\\
$c$ & 0.0069\\
$e$ & 0.00139 GeV$^{-1}$\\
$ N_0$ & $
         \begin{array}{lr}
         1.95\times10^{17}&  \mbox{for}\;\;\nu \\
         1.35\times10^{17}&  \;\mbox{for}\;\;\overline\nu.
         \end{array}$
         \\
         \hline
         \hline
         \end{tabular}
\caption{Parameters for the atmospheric $\nu_\mu+\bar{\nu}_\mu$ flux, in units of
$\GeV^{-1}{\rm km}^{-2}{\rm yr}^{-1}{\rm sr}^{-1}$.}
\label{table:atm}
\end{table}

\begin{figure}[h]
\begin{center}
\epsfig{file=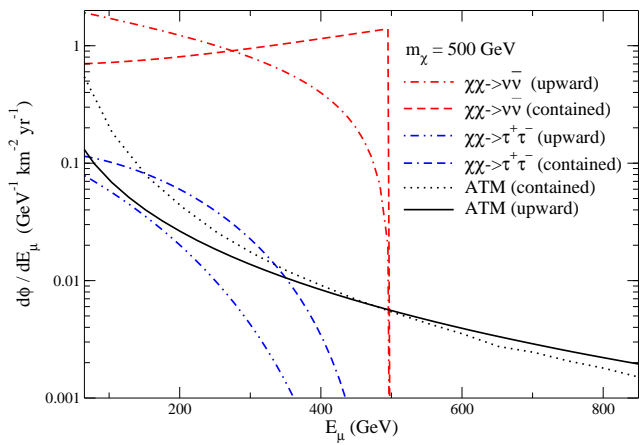,width=2.9in,angle=270}
\end{center}
\caption{Muon flux obtained from dark matter annihilation into 
neutrinos in the 
core of the Earth, 
 when muons are created in neutrino interactions with nucleons in 
the rock below the detector 
(dot-dashed and dot-dot-dashed curves), when muons are created in the detector, i.e. 
contained events (dashed and dot-dash-dashed curves).  The upper 
curves are for the direct production of 
neutrinos, while the lower 
curves are for neutrinos from tau 
decays.  The background from contained atmospheric neutrinos, evaluated for a cone of angle 
$\theta=1^\circ$ are shown with the dotted (black)
curve 
and the upward 
muon flux from atmospheric neutrinos is shown by the solid (black) curve.}
\label{fig:earth}
\end{figure}

\begin{figure}[h]
\begin{center}
\epsfig{file=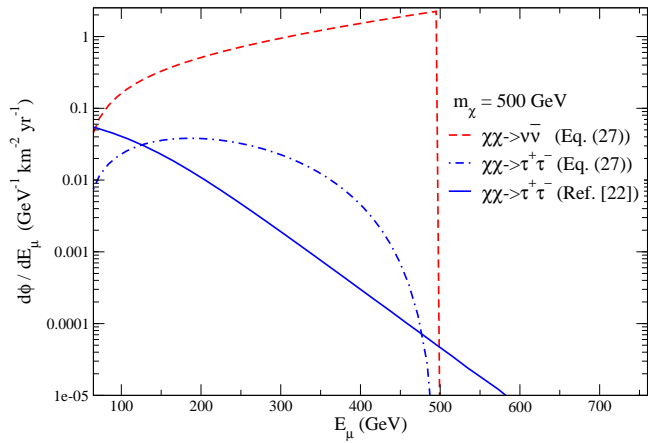,width=2.9in,angle=270}
\end{center}
\caption{Upward muons flux obtained using Eq. (27) for 
 $\chi \chi\rightarrow
\nu_\mu\bar{\nu}_\mu$ 
(dashed curve) and for 
$\chi\chi\rightarrow \tau^+\tau^-$, followed by $\tau\rightarrow
\nu_\tau \mu\bar{\nu}_\mu$ 
(dot-dashed curve), and the muon upward flux for 
$\chi\chi\rightarrow \tau^+\tau^-$ channel from Ref. \cite{edsjo} 
(solid 
curve). The upward muon flux from Eq. (27) is inconsistent with the upward flux shown in Fig. 1.} 
\label{fig:earthcompare}
\end{figure}

The energy dependence of Fig. \ref{fig:earth} is at odds with the muon energy
dependence sometimes found in the literature \cite{barger,WIMP,perez,menon}. There,
the upward flux of muons is written as 
\begin{eqnarray}\label{wrong}
\frac{d\phi_\mu}{dE_\mu} &=& \frac{\Gamma_A}{4\pi{R_E^2}}
\int^{m_\chi}_{E_\mu}dE_\nu\left(\frac{dN_\nu}{dE_\nu}\right)_{F,\mu} 
R_\mu(E_\mu,E_{th})\nonumber\\
& &\times\left\{\frac{d\sigma^p_\nu}{dE_\mu}\rho_p+(p\rightarrow{n})\right\}+(\nu\rightarrow\overline{\nu}),
\end{eqnarray}
where $E_{th}=50$ GeV. This expression accounts for the fact that muons have a range with an energy dependence,
however, it does not account for the fact
that over the distance $R(E_\mu,E_{th})$, the muon has a final energy of $E_{th}$.
Eq. (\ref{wrong}) 
does not represent the energy dependent muon flux, however, the integral number of 
upward events with $E_\mu>E_{th}$ obtained using Eq. (\ref{wrong}) and the results using 
Eq. (22) are approximately equal.  In Fig. \ref{fig:earthcompare}, we show the 
upward muon fluxes from 
Eq. (\ref{wrong}), for the direct neutrino production (dashed curve) and from the 
$\tau$ decay (dot-dashed curve).   
Comparing results from Figs. 1 and 2, we find that 
the upward muon flux of Eq. (\ref{wrong}) for 
$\chi \chi \rightarrow \nu \bar{\nu}$ case follows more closely the contained muon 
flux at high energies presented in Fig.1 (dashed curve) than the upward flux, with an enhancement at 
high $E_\mu$ because the muon range increases with muon energy.  Clearly, the upward muon flux in 
Fig. 2  (dashed curve) does not accurately reflect the muon energy distribution of upward events from 
DM annihilation in the Earth.  
Similarly, a comparison of upward muon flux for 
$\chi\chi\rightarrow \tau^+\tau^-$, followed by $\tau\rightarrow
\nu_\tau \mu\bar{\nu}_\mu$, obtained using Eq. (\ref{wrong}) has a very different shape than 
the same flux obtained with Eq. (23). 
Comparable discrepancies are found between upward fluxes from Eq. (\ref{wrong}) and our 
evaluation of upward events for DM annihilation in the Sun as well.

We also show with the solid line in Fig. \ref{fig:earthcompare}
the results for upward muon flux from the $\chi\chi\rightarrow \tau^+\tau^-$
from Ref. \cite{edsjo}.
In Ref. \cite{edsjo}, the flux of muons comes from a 
PYTHIA simulation of the resultant muon neutrino flux and a simulation of
muon electromagnetic energy loss. A 
dark matter distribution has been assumed in the Earth's core and 
contribution from dark matter annihilation around the center of the core 
with specific angular cuts ($\theta \leq 5^\circ$) have 
been applied, so the normalization should be lower.  
The energy distribution has qualitatively the same behavior as our results, however, 
it does not vanish at the kinematic limit when $E_\mu=m_\chi$.  

\subsection{DM annihilation in the Sun}

Similar conclusions can be derived in the case of capture of WIMPs 
in the core of the Sun.  As noted earlier, there is 
attenuation of the initial neutrino flux as it propagates from the core to the exterior of 
the Sun.  The interaction length of the neutrinos with energy $\sim30$ GeV 
becomes 
equal to the column depth of the Sun (the average density of the core of the Sun is 
$\sim 150$ 
g/cm$^3$).  
At higher energies, the interaction length becomes even smaller and 
the neutrino flux is reduced significantly.  
We do not include neutrino oscillation 
in the Sun \cite{barger}, which depending on the dark matter model, might affect the flux 
of $\nu_\mu+\bar{\nu}_\mu$.  

In Fig. 3, we show the upward muon and the 
contained muon fluxes for the direct production and for the 
$\tau$ production channels. In our calculations, we approximate neutrino attenuation 
in the Sun with 
an exponential suppression  
as presented in the previous section. We note that this effect becomes 
stronger for higher neutrino energies which manifests itself when $m_\chi$ is large. 
Recall that the charged current neutrino nucleon cross section  
increases with the neutrino energy. As an example,
the muon flux decreases by a 
factor of $3$ for $m_\chi=250\ \GeV$, factor of $10$ for $m_\chi=500$ GeV 
and two orders of magnitude
 for $m_\chi= 1$ TeV, as compared to the case with no attenuation.  

\begin{figure}[h]
\begin{center}
\epsfig{file=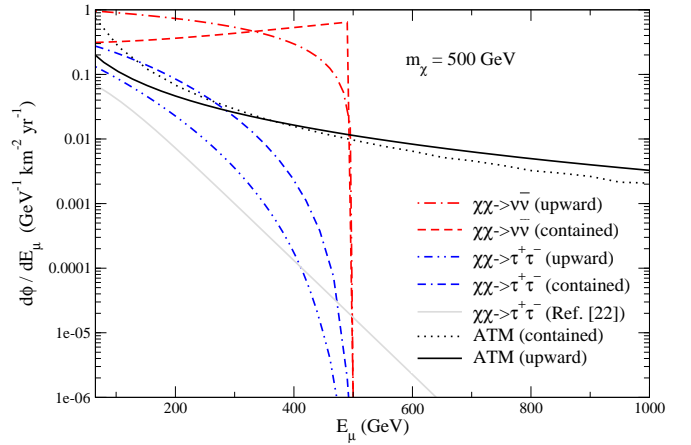,width=2.9in,angle=270}
\label{sun} 
\end{center}
\caption{Muon fluxes obtained from dark matter annihilation into 
neutrinos in the 
core of the Sun, 
for upward events (dot-dashed and dot-dot-dashed curves), and for 
contained events (dashed and dot-dash-dashed curves).  
The upper 
curves are for the direct production of 
neutrinos, while the lower 
curves are for neutrinos from tau 
decays.  Background upward muons are shown with the solid (black) curve and 
the contained muons are shown with the 
dotted (black) curve, where the evaluation used the angle-averaged atmospheric neutrino flux
integrated over a solid angle with
$\theta=1^\circ$. The grey solid curve is from Edsj\"o's parameterization of the muon flux \cite{edsjo}.}
\label{fig:sun}
\end{figure}

We compare our results for muon flux with those in Ref. \cite{edsjo}, where 
there is assumption of dark matter distribution in the core of the Sun and
contribution from dark matter annihilation around the center of the core
with specific angular cuts have been applied.   
Effects due to neutrino flavor oscillations in the Sun 
have not been incorporated.  The shape of the energy distribution is 
similar to our result, but with lower normalization and with a lack of 
the kinematic cutoff when $E_\mu=m_\chi$.  

As in the case of the Earth, the upward muon flux from 
$\chi \chi \rightarrow \nu \bar{\nu}$ is larger than the 
contained flux for muon energies, $E_\mu < 380$ GeV, while in the 
case when neutrinos are produced via 
$\chi\chi\rightarrow \tau^+\tau^-$, followed by $\tau\rightarrow
\nu_\tau \mu\bar{\nu}_\mu$, the contained muon flux is always larger than the upward flux.  
We also show the angle-averaged atmospheric flux for a cone of half-angle $1^\circ$. For direct
annihilation into neutrinos for the model in which the branching fraction is of the order of one, 
the signal is larger than the atmospheric background for both contained and upward muons. For the
tau channel, signal is comparable to the background for upward muons when muons have energy around 
$200$ GeV, however taking into account the effects of kinematics on the angular pointing of the muons at low energy 
may make this less apparent.  
 
\subsection{Muons in IceCube}

With the upward muon fluxes evaluated above from annihilation of DM in the Earth and the Sun, it is possible to estimate the event rate of muons in IceCube using the muon effective area \cite{gghm}. Following Ref. \cite{gghm}, we parameterized
\begin{equation}
A_{eff}(E_\mu,\theta) \simeq 2\pi A_0(E_\mu)(0.92-0.45\cos\theta )
\end{equation}
where $\theta$ is the zenith angle measured from vertical and 
\begin{eqnarray}
\nonumber
E_\mu &\leq& 10^{1.6}\ \GeV: \\ \nonumber
A_0(E_\mu)&=&0\\ \nonumber
10^{1.6}\ \GeV&\leq & E_\mu\leq 10^{2.8}\ \GeV:\\ \nonumber
A_0(E_\mu)&=& 0.748(\log_{10}(E_\mu/\GeV) - 1.6)\ {\rm km}
\\ \nonumber
10^{2.8}\ \GeV&\leq &E_\mu :\\ \nonumber
A_0(E_\mu)&=& 0.9+0.54(\log_{10}(E_\mu/\GeV) - 2.8)\ {\rm km}\ .
\end{eqnarray}
This effective muon area models the threshold detection effects near $E_\mu\sim 50$ GeV and local rock and ice below the IceCube detector
\cite{gghm}.

To facilitate comparisons with other muon energy distributions which appear in the literature, we evaluate
\begin{equation}
\label{eq:dNdE}
\frac{dN_\mu}{dE_\mu} = \frac{d\phi_\mu}{dE_\mu}\cdot\langle A_{eff}(
E_\mu,\theta )\rangle
\end{equation}
for DM annihilation to neutrinos in the Earth and Sun which convert to
muons outside the detector. Here,
$\langle A_{eff}\rangle$ is the angle averaged effective area, averaged over zenith angles $\theta = \pi/2  -\pi$.
Fig. \ref{fig:phiA} shows our results for the upward muon flux times effective area with the solid and dot-dashed lines (solid for the Earth), and by 
comparison, the 
results for the contained muon flux multiplied by 1 km$^2$ (dotted and dashed lines, dotted for the Earth). The energy dependence of the effective area 
changes 
the shapes of the curves for upward muons at low energies, but it does not change the large discrepancies between our upward muon rates compared with Eq. (\ref{wrong}) 
at energies closer to $E_\mu\sim m_\chi$.

\begin{figure}[h]
\begin{center}
\epsfig{file=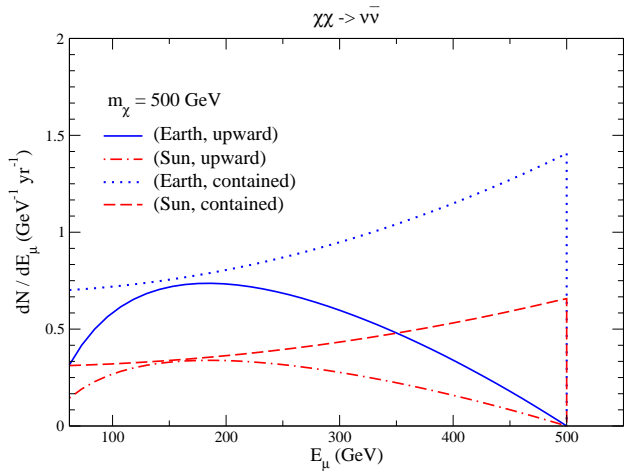,width=2.9in,angle=270}
\end{center}
\caption{The upward muon flux times muon effective area obtained from dark matter annihilation to neutrinos in the core of the Earth (solid line) and 
Sun (dot-dashed line). For comparison, we also show the contained muon flux times 1 km$^2$ for the Earth (dotted) and for the Sun (dashed).} 
\label{fig:phiA}
\end{figure}

\subsection{Cosmic Diffuse Neutrino Flux} 

In addition to the astrophysical object such as the 
Sun and the Earth being potential sources of dark matter, relic dark matter 
can also annihilate in 
halos in the universe \cite{beacom}, providing a promising source of 
cosmic diffuse neutrinos from dark matter annihilation.  

To determine this flux one needs to sum over all halos to yield a 
flux of neutrinos. 
This diffuse neutrino flux depends 
on several factors such as the evolution 
with redshift, the radial density 
profiles and the number density of halos of a given mass at a 
given redshift  
 \cite{beacom,ullio,ando}.  
 In Ref. \cite{beacom},
dark matter annihilation process,  
$\chi\chi\rightarrow\nu\overline\nu$, 
is proposed to be used to determine an upper limit on the annihilation cross section.  

The cosmic diffuse neutrinos for the 
$\chi\chi\rightarrow\nu\overline\nu$ channel 
from Ref. \cite{beacom} 
 is
approximately a power law function of $E_\nu$, i.e. 
\begin{equation}
\left(\frac{d\phi_\nu}{dE_\nu\,d\Omega}\right)_{\nu_\mu+\overline{\nu}_\mu}\simeq
A\frac{(E_\nu/\GeV)^{0.5}}
{(m_\chi/\GeV)^{3.5}}\quad \,\,\,\,  
 E_\nu\leq m_\chi \ .
\label{eq:diffuse}
\end{equation}
In Ref. \cite{beacom}, the normalization $A$ is determined by setting  
 the number of neutrinos from the diffuse flux (here approximated by Eq. (\ref{eq:diffuse})) 
equal to the number of atmospheric neutrinos from the same energy interval, from
$10^{-0.5}m_\chi$ to $m_\chi$, i.e. 

\begin{equation}
\int _{\frac{m_\chi}{\sqrt{10}}}^{m_\chi}
d E_\nu A\frac{(E_\nu/\GeV)^{0.5}}
{(m_\chi/\GeV)^{3.5}} 
= \int _{\frac{m_\chi}{\sqrt{10}}}^{m_\chi}
d E_\nu \left(\frac{d\phi_\nu}{dE_\nu{d\Omega}}\right)_{av} 
\end{equation}
where $\left(\frac{d\phi_\nu}{dE_\nu{d\Omega}}\right)_{av}$ is the 
 angle-averaged atmospheric flux 
given by Eq. (26).  We consider here a case when 
 $m_\chi=1$ TeV.  

In Fig.
\ref{fig:diffuse}, we show upward and contained muon fluxes from cosmic diffuse neutrinos, integrating over
the full $2\pi$ solid angle. For $E_\mu>400$ GeV, the contained flux
dominates the upward flux, an artifact of the triangular shape of the neutrino flux.
For comparison we also show the contained and upward fluxes for the atmospheric neutrino background.
The falling energy spectrum of the atmospheric neutrinos results in the upward flux of muons from atmospheric neutrinos dominating the contained muon flux for $E_\mu>  400$ GeV for $m_\chi=1$ TeV.

\begin{figure}[h]
\begin{center}
\epsfig{file=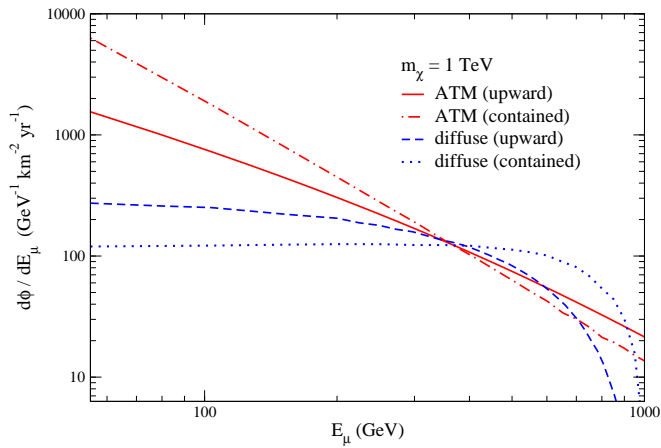,width=2.9in,angle=270}
\end{center}
\caption{Muon fluxes obtained from dark matter annihilation in the halos producing 
cosmic diffuse neutrinos: upward muons flux 
(dashed curve) and 
 contained muon flux 
 (dotted curve) compared with 
 contained 
 (dot-dashed curve) and upward 
 (solid curve) muon fluxes 
from angle-averaged atmospheric neutrinos. Here, we take
$m_\chi = 1$ TeV.} 
\label{fig:diffuse}
\end{figure}

The direct production of neutrinos
$\chi\chi\rightarrow \nu \bar{\nu}$ is the most favorable channel in terms of neutrino detection
for the diffuse 
DM limits since the muon flux stands out more from the background than the muons 
from a $\chi\rightarrow \tau\rightarrow\nu_\mu$ cascade 
(or similar production and decay process).  In addition, the $\chi\chi\rightarrow \nu\bar{\nu}$ channel
has no other astrophysical observable.
Nevertheless, the muon flux is not as dramatic a peak in the 
falling neutrino induced atmospheric muon flux as the direct comparison of the neutrino fluxes is.
A more comprehensive analysis of a diffuse DM annihilation signal could include 
both the $\nu\bar{\nu}$ and cascade channels as possibilities, 
and focus on the muon signals rather than the neutrino signals.

\section{Conclusions}
We have calculated muon fluxes from dark matter annihilation,
when dark matter is trapped in the the Sun's (Earth's) core and when
dark matter annihilates in halos in the universe (cosmic diffuse flux).
Without using a specific model for dark matter, we have considered
 $\chi\chi\rightarrow \nu\bar{\nu}$ and
$\chi\chi\rightarrow \tau^+\tau^-$, followed by $\tau\rightarrow
\nu_\tau \mu\bar{\nu}_\mu$ channels as  
 representatives of direct and of the secondary neutrino production.
We have taken into account neutrino attenuation as it propagates from
the core of the Sun to its surface.
In the evaluation of the upward muon flux, we have incorporated muon
energy loss, as described by the muon range.

 We have shown that
our results exhibit a very different energy dependence than
those obtained from Eq. (\ref{wrong}) that is widely used 
in the literature \cite{barger,perez,menon}, however, there is reasonably 
good agreement with the
parameterization of Ref. \cite{edsjo} away from the region of maximum energy 
$E_\mu\sim m_\chi$.
Our results are obtained with the
assumption that the dark matter annihilation occurs at the maximum rate, 
when
 the
annihilation rate is half the capture rate.  This is reasonable 
 for the Sun but requires significant
enhancement of the capture rate (or annihilation cross section) for the
Earth to be in equilibrium \cite{perez}.   

In our calculation we used 
spin independent WIMP-nucleon cross sections which have much stronger 
experimental bound than the 
spin dependent cross sections \cite{SD}.  In 
 the core of the Sun the capture rate 
might be dominated by the 
spin dependent (SD) WIMP-hydrogen nuclei interactions, 
which would increase the signal 
rates by a couple of orders of magnitude and still be consistent with 
Amanda limits on annihilation rates \cite{Amanda_limits}.  
In the dark matter model in which there is a 
 low velocity enhancement of the DM annihilation cross section \cite{arkani}, 
 introduced 
 as an explanation for the positron 
excess observed in cosmic ray experiments \cite{HEAT,PAMELA,ATIC,PPB-BETS}, 
it is possible for the WIMPs in the core of the Earth to be in the equilibrium as well.  

Furthermore,
 incorporating neutrino oscillations and the regeneration effects in the 
Sun
will likely affect the final muon flux especially in the models 
which possess
an asymmetry in the initial neutrino fluxes or where
$\chi\chi\rightarrow\tau^+\tau^-$ is the
dominant mode \cite{barger}.   We have used a model independent  
normalization,
$\sigma_0^i\simeq10^{-8}N^4_i$ pb and $B_F=1$ to
evaluate the muon flux.  We find that for this branching fraction
signals from
 $\chi\chi\rightarrow \nu\bar{\nu}$ and
$\chi\chi\rightarrow \tau^+\tau^-$, followed by $\tau\rightarrow
\nu_\tau \mu\bar{\nu}_\mu$, when DM annihilation happens in the
core of the Sun, are comparable or even larger than the
background (upward) muons from atmospheric neutrinos.
In the case of direct neutrino production, the upward muon flux is    
larger than the contained flux for $E_\mu < 350 $ GeV for $m_\chi=500$ GeV, due to
the muon range.  When neutrinos are produced via secondary processes,
contained events always dominate upward muons.

Cosmic diffuse neutrinos, produced directly in DM annihilation,
give a very weak energy dependence of the contained and upward muon
flux, in contrast to the steep energy dependence of the atmospheric
background.   The upward muon flux from cosmic diffuse neutrinos is
dominant over the contained flux for muon energies below $400$ GeV for $m_\chi=1$ TeV,
which also happens to be the energy at which the signal becomes
dominant over the background muons from
atmospheric neutrinos.

Model dependence is an important element, for
example, $\chi\chi\rightarrow \nu\bar{\nu}$ is not allowed for DM at rest when the DM particles are neutralinos \cite{barger}.
However,
with the formalism developed,
one can determine
 muon fluxes for specific dark matter model by
summing up
the contributions from all decay channels weighted with
corresponding branching fractions \cite{erkoca}.
Thus, measurements of the muon energy distribution in neutrino telescopes, such as IceCUBE 
and KM3, could provide valuable information about the origin of 
the dark matter sector and fundamental properties such as the 
dark matter mass and its couplings.  
\begin{acknowledgments}
M.H.R. and I.S.would like to thank Aspen Center for Physics for 
hospitality while this work was completed. We thank D. Marfatia and V.
Barger for discussions.
This research was supported by
US Department of Energy
contracts DE-FG02-91ER40664, DE-FG02-04ER41319 and DE-FG02-04ER41298.
\end{acknowledgments}

\newpage
$$ $$ 
\vskip 0.1true in
\section{Appendix:Muon Neutrino Distributions}

\subsection{Neutrino energy distribution from direct production:} 

Neutrino energy distribution when neutrinos are produced directly from 
dark matter annihilation is given by, 

\begin{equation}
\frac{dN_\nu}{dE_\nu}=\delta(E_\nu-m_\chi)
\end{equation}

\subsection{Neutrino energy distribution from $\tau^+\tau^-, b\overline{b}, c\overline{c}$ decay 
modes:}

In these decay modes, we use the 
unpolarized decay distributions, so the 
$\nu$ and $\overline\nu$ distributions are assumed to be 
the same.  
The decay 
branching fraction is denoted by $B_f$ for a given decay mode $f$, 
$f=\nu,\tau,b,c$. The $b$ and $c$ quarks hadronize before they decay into 
neutrinos. The hadronization effect is taken into account by scaling 
the initial quark energy, $E_{in}$, in the form $E_d=z_fE_{in}$,
where $z_f=0.73,0.58$ for $b$ and $c$ quarks, respectively \cite{jungman2}. 

Neutrino energy distribution from the decay of 
$\tau^+\tau^-, b\overline{b}, c\overline{c}$ is approximately 
\begin{equation}
\frac{dN_\nu}{dE_\nu}=\frac{2B_f}{E_{in}}(1-3x^2+2x^3),\;\;\; \mbox{where}\;\;\; 
x=\frac{E_\nu}{E_{in}}\le1\ ,
\end{equation}   
where 
   \begin{equation}
(E_{in}\;,\;B_f)=\left\{
   \begin{array}{lr}
   (m_\chi\;,\;0.18) &           \tau \;\;\mbox{decay}, \\
   (0.73m_\chi\;,\;0.103) &       b\;\;  \mbox{decay}, \\
   (0.58m_\chi\;,\;0.13)&       c\;\;  \mbox{decay}.
   \end{array}  
   \right.
\end{equation}

\end{document}